\input harvmac.tex


\input epsf.tex
\def\figin{\epsfcheck\figin}\def\figins{\epsfcheck\figins}
\def\epsfcheck{\ifx\epsfbox\UnDeFiNeD
\message{(NO epsf.tex, FIGURES WILL BE IGNORED)}
\gdef\figin##1{\vskip2in}\gdef\figins##1{\hskip.5in}
\else\message{(FIGURES WILL BE INCLUDED)}%
\gdef\figin##1{##1}\gdef\figins##1{##1}\fi}
\def\DefWarn#1{}
\def\figinsert{\goodbreak\midinsert}
\def\ifig#1#2#3{\DefWarn#1\xdef#1{fig.~\the\figno}
\writedef{#1\leftbracket fig.\noexpand~\the\figno}%
\figinsert\figin{\centerline{#3}}\medskip\centerline{\vbox{\baselineskip12pt
\advance\hsize by -1truein\noindent\footnotefont{\bf
Fig.~\the\figno:} #2}}
\bigskip\endinsert\global\advance\figno by1}

\def\ndt{\noindent}

\def\K3{{\bf K3}}
\def\journal#1&#2(#3){\unskip, \sl #1\ \bf #2 \rm(19#3) }
\def\andjournal#1&#2(#3){\sl #1~\bf #2 \rm (19#3) }

\def\bar{\overline}
\def\hat{\widehat}

\def\tilde{\widetilde}

\def\frac#1#2{{#1\over#2}}

\def\inbar{\,\vrule height1.5ex width.4pt depth0pt}
\def\IC{\relax\hbox{$\inbar\kern-.3em{\rm C}$}}
\def\IR{\relax{\rm I\kern-.18em R}}
\def\IP{\relax{\rm I\kern-.18em P}}

%
%


%
\catcode`\@=11
\def\slash#1{\mathord{\mathpalette\c@ncel{#1}}}
\overfullrule=0pt

\def\CC{{\cal C}}

\def\LL{{\cal L}}
\def\NN{{\cal N}}

\def\lam{\lambda}

\def\underrel#1\over#2{\mathrel{\mathop{\kern\z@#1}\limits_{#2}}}

\catcode`\@=12


%

\def\det{{\rm det}}

\def\det{{\rm det}}
\def\exp{{\rm exp}}


\def \ov {\over}
\def \p {\partial}
\def \ha {{1 \ov 2}}
\def \al {\alpha}
\def \lam {\lambda}

\def \sig {\sigma}

\def \om {\omega}
\def \Om {\Omega}
\def \ep {\epsilon}

\def \ga {\gamma}
\def \apr {\alpha'}

\def\le{\left}
\def\ri{\right}

\def\th{\theta}

\def\IL{\relax{\rm I\kern-.18em L}}
\def\IH{\relax{\rm I\kern-.18em H}}
\def\IR{\relax{\rm I\kern-.18em R}}
\def\IC{\relax\hbox{$\inbar\kern-.3em{\rm C}$}}
\def\IZ{{\bf Z}}





\def\makeblankbox#1#2{\hbox{\lower\dp0\vbox{\hidehrule{#1}{#2}%
   \kern -#1
   \hbox to \wd0{\hidevrule{#1}{#2}%
      \raise\ht0\vbox to #1{}
      \lower\dp0\vtop to #1{}
      \hfil\hidevrule{#2}{#1}}%
   \kern-#1\hidehrule{#2}{#1}}}%
}%
\def\hidehrule#1#2{\kern-#1\hrule height#1 depth#2 \kern-#2}%
\def\hidevrule#1#2{\kern-#1{\dimen0=#1\advance\dimen0 by #2\vrule
    width\dimen0}\kern-#2}%
\def\openbox{\ht0=1.2mm \dp0=1.2mm \wd0=2.4mm  \raise 2.75pt
\makeblankbox {.25pt} {.25pt}  }

\def\bun#1/#2{\leavevmode
   \kern.1em \raise .5ex \hbox{\the\scriptfont0 #1}%
   \kern-.1em $/$%
   \kern-.15em \lower .25ex \hbox{\the\scriptfont0 #2}%
}

\def\opensquare{\ht0=3.4mm \dp0=3.4mm \wd0=6.8mm  \raise 2.7pt
\makeblankbox {.25pt} {.25pt}  }


\def\sector#1#2{\ {\scriptstyle #1}\hskip 1mm
\mathop{\opensquare}\limits_{\lower
1mm\hbox{$\scriptstyle#2$}}\hskip 1mm}

\def\tsector#1#2{\ {\scriptstyle #1}\hskip 1mm
\mathop{\opensquare}\limits_{\lower
1mm\hbox{$\scriptstyle#2$}}^\sim\hskip 1mm}

\def\IZ{{\bf Z}}


\lref\LV{Landsman and Van Weert , Phys.\ Rept.\  }

\lref\CrnkovicMS{ C.~Crnkovic, M.~R.~Douglas and G.~W.~Moore,
``Physical Solutions For Unitary Matrix Models,'' Nucl.\ Phys.\ B
{\bf 360}, 507 (1991).
}

\lref\SchnitzerQT{ H.~J.~Schnitzer, ``Confinement / deconfinement
transition of large N gauge theories with N(f) fundamentals:
N(f)/N finite,'' hep-th/0402219.
}

\lref\aharony{ O.~Aharony, J.~Marsano, S.~Minwalla, K.~Papadodimas
and M.~Van Raamsdonk, ``The Hagedorn / deconfinement phase
transition in weakly coupled large N gauge theories,''
hep-th/0310285.
}

\lref\sundborg{ B.~Sundborg, ``The Hagedorn transition,
deconfinement and N = 4 SYM theory,'' Nucl.\ Phys.\ B {\bf 573},
349 (2000) [hep-th/9908001];
 B.~Sundborg,
``Stringy gravity, interacting tensionless strings and massless
higher spins,'' Nucl.\ Phys.\ Proc.\ Suppl.\  {\bf 102}, 113
(2001) [hep-th/0103247].
}

\lref\gw{ D.~J.~Gross and E.~Witten, ``Possible Third Order Phase
Transition In The Large N Lattice Gauge Theory,'' Phys.\ Rev.\ D
{\bf 21}, 446 (1980).
}

\lref\wadia{ S.~R.~Wadia, ``N = Infinity Phase Transition In A
Class Of Exactly Soluble Model Lattice Gauge Theories,'' Phys.\
Lett.\ B {\bf 93}, 403 (1980).
}

\lref\malda{ J.~M.~Maldacena, ``The large N limit of
superconformal field theories and supergravity,'' Adv.\ Theor.\
Math.\ Phys.\  {\bf 2}, 231 (1998) [Int.\ J.\ Theor.\ Phys.\  {\bf
38}, 1113 (1999)] [hep-th/9711200].
}

\lref\witten{ E.~Witten, ``Anti-de Sitter space, thermal phase
transition, and confinement in  gauge theories,'' Adv.\ Theor.\
Math.\ Phys.\  {\bf 2}, 505 (1998) [arXiv:hep-th/9803131].
}

\lref\hawkingpage{ S.~W.~Hawking and D.~N.~Page, ``Thermodynamics
Of Black Holes In Anti-De Sitter Space,'' Commun.\ Math.\ Phys.\
{\bf 87}, 577 (1983).
}

\lref\atic{ J.~J.~Atick and E.~Witten, ``The Hagedorn Transition
And The Number Of Degrees Of Freedom Of String Theory,'' Nucl.\
Phys.\ B {\bf 310}, 291 (1988).
}

\lref\sapa{ B.~Sathiapalan, ``Vortices On The String World Sheet
And Constraints On Toral Compactification,'' Phys.\ Rev.\ D {\bf
35}, 3277 (1987).
}

\lref\kogan{ Y.~I.~Kogan, ``Vortices On The World Sheet And
String's Critical Dynamics,'' JETP Lett.\  {\bf 45}, 709 (1987)
[Pisma Zh.\ Eksp.\ Teor.\ Fiz.\ {\bf 45}, 556 (1987)].
}

\lref\goldsm{ Y.~Y.~Goldschmidt, ``1/N Expansion In
Two-Dimensional Lattice Gauge Theory,'' J.\ Math.\ Phys.\  {\bf
21}, 1842 (1980).
}

\lref\polaP{ J.~Jurkiewicz and K.~Zalewski, ``Vacuum Structure Of
The U(N $\to$ Infinity) Gauge Theory On A Two-Dimensional Lattice
For A Broad Class Of Variant Actions,'' Nucl.\ Phys.\ B {\bf 220},
167 (1983).
}

\lref\periw{ V.~Periwal and D.~Shevitz, ``Unitary Matrix Models As
Exactly Solvable String Theories,'' Phys.\ Rev.\ Lett.\  {\bf 64},
1326 (1990); \quad

 V.~Periwal and D.~Shevitz, ``Exactly Solvable Unitary
Matrix Models: Multicritical Potentials And
Nucl.\ Phys.\ B {\bf 344}, 731 (1990).
}

\lref\kms{ I.~R.~Klebanov, J.~Maldacena and N.~Seiberg, ``Unitary
and complex matrix models as 1-d type 0 strings,'' hep-th/0309168.
}

\lref\klebanov{ I.~R.~Klebanov, ``Touching random surfaces and
Liouville gravity,'' Phys.\ Rev.\ D {\bf 51}, 1836 (1995)
[hep-th/9407167].
}

\lref\klebC{ I.~R.~Klebanov and A.~Hashimoto, ``Nonperturbative
solution of matrix models modified by trace squared terms,''
Nucl.\ Phys.\ B {\bf 434}, 264 (1995) [hep-th/9409064]; \quad

J.~L.~F.~Barbon, K.~Demeterfi, I.~R.~Klebanov and C.~Schmidhuber,
``Correlation functions in matrix models modified by wormhole
terms,'' Nucl.\ Phys.\ B {\bf 440}, 189 (1995) [hep-th/9501058].
}

\lref\das{ S.~R.~Das, A.~Dhar, A.~M.~Sengupta and S.~R.~Wadia,
``New Critical Behavior In D = 0 Large N Matrix Models,'' Mod.\
Phys.\ Lett.\ A {\bf 5}, 1041 (1990).
}

\lref\BranV{ R.~H.~Brandenberger and C.~Vafa, ``Superstrings In
The Early Universe,'' Nucl.\ Phys.\ B {\bf 316}, 391 (1989).
}

\lref\TanI{ N.~Deo, S.~Jain, O.~Narayan and C.~I.~Tan, ``The
Effect of topology on the thermodynamic limit for a string gas,''
Phys.\ Rev.\ D {\bf 45}, 3641 (1992).
}

\lref\polyakov{ A.~M.~Polyakov, ``Gauge fields and space-time,''
Int.\ J.\ Mod.\ Phys.\ A {\bf 17S1}, 119 (2002)
[arXiv:hep-th/0110196].
}

\lref\bipz{E.~Brezin, C.~Itzykson, G.~Parisi and J.~B.~Zuber,
``Planar Diagrams,'' Commun.\ Math.\ Phys.\  {\bf 59}, 35 (1978).
 }

\lref\HallinKM{
J.~Hallin and D.~Persson,
``Thermal phase transition in weakly interacting, large N(c) {QCD},''
Phys.\ Lett.\ B {\bf 429}, 232 (1998)
[arXiv:hep-ph/9803234].
}

\lref\FuruuchiSY{
K.~Furuuchi, E.~Schreiber and G.~W.~Semenoff,
``Five-brane thermodynamics from the matrix model,''
arXiv:hep-th/0310286;
\quad G.~W.~Semenoff, ``Matrix model thermodynamics,''
arXiv:hep-th/0405107.
}

\lref\GubserBC{
S.~S.~Gubser, I.~R.~Klebanov and A.~M.~Polyakov,
``Gauge theory correlators from non-critical string theory,''
Phys.\ Lett.\ B {\bf 428}, 105 (1998)
[arXiv:hep-th/9802109].
}

\lref\WittenQJ{
E.~Witten,
``Anti-de Sitter space and holography,''
Adv.\ Theor.\ Math.\ Phys.\  {\bf 2}, 253 (1998)
[arXiv:hep-th/9802150].
}

\lref\AharonyIG{
O.~Aharony, J.~Marsano, S.~Minwalla and T.~Wiseman,
``Black hole - black string phase transitions in thermal 1+1 dimensional
arXiv:hep-th/0406210.
}

\lref\FidkowskiFC{
L.~Fidkowski and S.~Shenker,
``D-brane instability as a large N phase transition,''
arXiv:hep-th/0406086.
}

\lref\HaggiManiRU{
P.~Haggi-Mani and B.~Sundborg,
``Free large N supersymmetric Yang-Mills theory as a string theory,''
JHEP {\bf 0004}, 031 (2000)
[arXiv:hep-th/0002189].
}

\lref\SezginZS{
E.~Sezgin and P.~Sundell,
``Doubletons and 5D higher spin gauge theory,''
JHEP {\bf 0109}, 036 (2001)
[arXiv:hep-th/0105001];

A.~Mikhailov,
``Notes on higher spin symmetries,''
arXiv:hep-th/0201019;

E.~Sezgin and P.~Sundell,
``Massless higher spins and holography,''
Nucl.\ Phys.\ B {\bf 644}, 303 (2002)
[Erratum-ibid.\ B {\bf 660}, 403 (2003)]
[arXiv:hep-th/0205131];

M.~Bianchi, J.~F.~Morales and H.~Samtleben,
``On stringy AdS(5) x S**5 and higher spin holography,''
JHEP {\bf 0307}, 062 (2003)
[arXiv:hep-th/0305052];

N.~Beisert, M.~Bianchi, J.~F.~Morales and H.~Samtleben,
``Higher spin symmetry and N = 4 SYM,''
arXiv:hep-th/0405057.
}

\lref\malet{
J.~M.~Maldacena,
``Eternal black holes in Anti-de-Sitter,''
JHEP {\bf 0304}, 021 (2003)
[arXiv:hep-th/0106112].
}

\lref\witt{
E.~Witten,
``Some comments on string dynamics,''
arXiv:hep-th/9507121.
}

\lref\vafao{
H.~Ooguri and C.~Vafa,
``Two-Dimensional Black Hole and Singularities of CY Manifolds,''
Nucl.\ Phys.\ B {\bf 463}, 55 (1996)
[arXiv:hep-th/9511164].
}

\lref\kuta{
A.~Giveon and D.~Kutasov,
``Little string theory in a double scaling limit,''
JHEP {\bf 9910}, 034 (1999)
[arXiv:hep-th/9909110].
}

\lref\AlvarezGaumeZI{ L.~Alvarez-Gaume, J.~L.~F.~Barbon and
C.~Crnkovic, ``A Proposal for strings at D > 1,'' Nucl.\ Phys.\ B
{\bf 394}, 383 (1993) [arXiv:hep-th/9208026].
}

\lref\KorchemskyTT{ G.~P.~Korchemsky, ``Matrix model perturbed by
higher order curvature terms,'' Mod.\ Phys.\ Lett.\ A {\bf 7},
3081 (1992) [arXiv:hep-th/9205014];
 G.~P.~Korchemsky, ``Loops in the curvature
matrix model,'' Phys.\ Lett.\ B {\bf 296}, 323 (1992)
[arXiv:hep-th/9206088].
}

\lref\BarbonDI{
J.~L.~F.~Barbon and E.~Rabinovici,
``Closed-string tachyons and the Hagedorn transition in AdS space,''
JHEP {\bf 0203}, 057 (2002)
[arXiv:hep-th/0112173];
J.~L.~F.~Barbon and E.~Rabinovici,
``Touring the Hagedorn Ridge,''
arXiv:hep-th/0407236.
}

\lref\GrossMR{
D.~J.~Gross and A.~Matytsin,
``Instanton induced large N phase transitions in two-dimensional and
Nucl.\ Phys.\ B {\bf 429}, 50 (1994)
[arXiv:hep-th/9404004].
}

\lref\ambj{
J.~Ambjorn, B.~Durhuus and T.~Jonsson,
``Quantum geometry. A statistical field theory approach,''
Cambridge Monogr.\ Math.\ Phys.\  {\bf 1}, 1 (1997).
}



\Title{\vbox{\baselineskip12pt \hbox{hep-th/0408001}
\hbox{MIT-CTP-3520}
}}%
{\vbox{\centerline{Fine structure of Hagedorn transitions}} }

\smallskip
\centerline{  Hong Liu\foot{hong\_liu@lns.mit.edu}}
\medskip

\centerline{\it  Center for Theoretical Physics} \centerline{\it
Massachusetts Institute of Technology} \centerline{\it Cambridge,
Massachusetts, 02139}

\smallskip

\smallskip

\smallskip

\vglue .3cm

\bigskip
\noindent

We study non-perturbative aspects of the Hagedorn transition for
IIB string theory in an anti-de Sitter spacetime in the limit that the
string length goes  to infinity. The theory has a holographic dual
in terms of free $\NN=4$ super-Yang-Mills theory on a
three-dimensional sphere. We define a double scaling limit in which
the width of the transition region around the Hagedorn temperature
scales with the effective string coupling with a critical exponent.
We show that in this limit the transition is smoothed out by quantum effects.
In particular, the Hagedorn singularity of perturbative string theory
is removed by summing over two different
string geometries: one from the thermal AdS background, the other
from a noncritical string
background. The associated noncritical string has the scaling of the
 unconventional branch of super-Liouville theory or a branched
polymer.

\Date{July 31, 2004}




\bigskip


\newsec{Introduction and summary}

One of the deep mysteries of perturbative string theory is the
nature of the Hagedorn transition. In all known string theories
with spacetime dimensions greater than two, there exists a
critical Hagedorn temperature, $T_H$, above which the thermal
partition function for free strings diverges. The divergence can
be attributed to the exponential growth of the number of
perturbative string states at high energies. It is generally
believed that rather than a limiting temperature for string
theory, the Hagedorn temperature signals a transition to a new
phase~\refs{\sapa,\kogan,\atic}, analogous to the deconfinement
transition in a nonabelian gauge theory. From the worldsheet point of
view, the Hagedorn temperature is associated with the appearance
of new relevant operators in the worldsheet conformal field
theory. String theory at a finite temperature can be described by
strings propagating in a Euclidean target space with time
direction $X^0$ periodically identified with a period given by the
inverse temperature. The winding modes in the Euclidean time
direction are irrelevant operators when the temperature $T$ is
small. The lowest winding  modes become marginal at $T_H$ and
become relevant above $T_H$. The emergence of these new relevant
operators signals that the worldsheet theory becomes unstable and
most likely will flow to a new infrared fixed point.

By studying the effective field theory near the Hagedorn temperature,
Atick and Witten~\refs{\atic} argued that the Hagedorn transition is a
first order transition that happens below the
Hagedorn temperature seen in perturbation theory. In particular,
the genus zero contribution to the free energy 
is nonzero above the transition, i.e. the free energy is of
order $g_s^{-2}$, where $g_s$ is the string coupling constant.
This implies that
one can no longer ignore the back reaction of the thermal energy
density on the background geometry, since no matter how small
$g_s$ is, the back reaction is always at least of order one.

One expects that a precise understanding of the Hagedorn
transition and physics of the high temperature phase should give
us important insights into the fundamental structure of string
theory. Given our very limited understanding of these important
questions, it is thus of great interest to find a simpler setting
that one could study similar questions in full detail.

Recently, it was found that free $\NN=4$ Yang-Mills theory with
gauge group $SU(N)$ on a three sphere $S^3$ also exhibits a
Hagedorn type transition in the limit $N \to
\infty$~\refs{\sundborg,\polyakov,\aharony}\foot{
See~\refs{\HallinKM,\FuruuchiSY,\SchnitzerQT,\FidkowskiFC,\AharonyIG}
for other recent discussions of phase transitions in weakly
coupled Yang-Mills theory.}. In the large $N$ limit the number of
gauge invariant operators grows exponentially with conformal
dimension. This leads to a Hagedorn temperature  $T_H$, given by
an order one constant times the curvature of $S^3$, above which
the thermal partition function diverges in the $N=\infty$ limit.
At large but finite $N$, one finds a weakly first order phase
transition in the $1/N$ perturbation theory, with the free energy
of order $O(1)$ in the low temperature phase, while of $O(N^2)$ in
the high temperature phase.

From the AdS/CFT correspondence~\refs{\malda,\GubserBC,\WittenQJ},
the $\NN=4$ SYM
theory with gauge group $SU(N)$ and coupling constant $g_{YM}$ on
$S^3$ describes type IIB superstring theory in $AdS_5 \times S_5$.
The string scale $l_s$ and  ten-dimensional Newton's constant $G$
are given from Yang-Mills parameters as\foot{We omit order one
numerical constants.}
$$
 { R^4 \ov l_s^4} = g_{YM}^2 N,
\qquad {G \ov R^8} 
= N^{-2} ,  \qquad G = g_s^2 l_s^8 = l_p^8,  \qquad g_s = g_{YM}^2
$$
where $R$ is the curvature radius of the AdS$_5$. The $1/N$
t'Hooft expansion in Yang-Mills theory corresponds to the $g_s$
expansion (quantum gravitational corrections) in AdS, and a small
t'Hooft coupling $g_{YM}^2 N$ implies a strongly coupled
worldsheet. Free $SU(N)$ Yang-Mills theory with large but finite
$N$ appears to describe a string theory in AdS$_5 \times S_5$ in
the limit
 \eqn\specL{
g_s \to 0, \quad l_s \to \infty, \quad {g_s^2 l_s^8 \ov R^8} = {\rm finite}
 \propto {1 \ov N^2}
 }
i.e. we are working in the regime
 \eqn\reime{
l_s = \infty \gg R \gg l_p \ .
 }
While very little is known about this
theory\foot{It has been conjectured that this theory might be related
to a theory of higher spins~\refs{\sundborg,\HaggiManiRU,\SezginZS}.},
if we assume the validity of the AdS/CFT correspondence in
this extreme case, we conclude that it
should have the following properties:

\item{$\bullet$} The theory has a perturbative stringy expansion in terms of
 genus expansion with an effective expansion parameter $G/R^8$.

 \item{$\bullet$} The theory has a Hagedorn divergence in perturbation
 theory due to exponential growth of perturbative stringy states.

 \item{$\bullet$} The free energy is $O(1)$ below $T_H$, while of order
 $G^{-1}$ above $T_H$.

\ndt Even though this AdS string theory%
\foot{See
also \BarbonDI\ for some recent discussions of Hagedorn transition
in large AdS.}
seems rather different from
our familiar flat space string theories, the features listed above
about the Hagedorn transition are qualitatively similar to those in
flat spacetime.
Here we
have the advantage  that the theory has a holographic description
in terms of a free Yang-Mills theory and thus its Hagedorn
transition can be studied in detail non-perturbatively.

The Hagedorn transition for free Yang-Mills theory is a large $N$
phase transition~\refs{\sundborg,\aharony}. The number of states
at a given energy $E$ can be enumerated by counting gauge
invariant operators of conformal dimension $E$. Since the theory
is free, this can be done exactly. At energies\foot{We will take
the radius of $S^3$ to be $1$.} $1 \ll E \ll N^2$, one can treat
all operators of given dimensions as independent. One finds that
the number of states grows as $e^{E \ov T_H}$ with $T_H$ given by
an order one constant. At energies $E \sim N^2$, the counting
becomes more complicated, since finite $N$ effects like trace
relations between single and multi-trace operators are important.
Treating all the gauge invariant operators as independent
dramatically overcounts the number of states, resulting in the
Hagedorn divergence. At energies $E \gg N^2$, instead of counting
gauge invariant operators, it is simpler to treat the system as
$N^2$ species of gluons and quarks. Then standard results for the
ideal gas implies that the number of states should be proportional
to $e^{c N^\ha E^{3 \ov 4}}$, which is much slower than the
Hagedorn growth. The Hagedorn transition can thus be understood as
the crossover from a region in which the gauge invariant operator
description is more appropriate to a region in which the
gluon-quark description is more appropriate. On a compact space
like $S^3$, a sharp transition can only happen in the strict
$N=\infty$ limit. At large but finite $N$, the transition arises
as an artifact of the large $N$ expansion\foot{More precisely, the
partition function is an analytic function of temperature $T$ for
all $N$. The non-analyticity in $T$ arises as a result of
expansion in $1/N$.}. The interpolation between the low and high
temperature regimes should be completely smooth at finite $N$ and
becomes narrower and sharper as $N$ is increased.

Translating the above physical picture to the corresponding string
theory in AdS, we conclude that

 \item{$\bullet$} The Hagedorn divergence in the bulk string theory
should be an artifact of the string perturbation theory.
The transition should be smooth non-perturbatively.

 \item{$\bullet$}
The
appearance of the Hagedorn divergence in
 perturbation theory can be attributed to
a ``stringy exclusion principle'', i.e.
not all perturbative states are independent of one another at high
energies.

\ndt In this paper we study non-perturbative aspects of this
Hagedorn transition using Yang-Mills theory. We would like to
understand how non-perturbative stringy effects in AdS smooth out
the sharp Hagedorn transition in perturbative string theory. In
other words, we are interested in knowing how the Hagedorn
divergence in free string theory is resolved at finite string
coupling~\foot{Naively  $g_{YM} =0$  seems to imply that string
coupling is also zero. The effective genus expansion parameter
$g_s^2 l_s^8/R^8$ in
\specL\ for the bulk string theory is nonzero at finite $N$.
 This is the sense
that we talk about the theory at finite string coupling throughout
the paper.}.

We now summarize the main results of the paper.
We will argue that the large-$N$ Hagedorn transition for free $\NN=4$
Super-Yang-Mills theory belongs to the same universality class as
that of the double-trace unitary matrix model~\refs{\aharony}
 \eqn\DEFMo{
 Z(a) = \int dU \, \exp \le[a (T) \Tr U \Tr U^\dagger \ri]
 }
where $U$ is a unitary matrix and  $a (T)$ is a function of
temperature. \DEFMo\ has a Hagedorn type transition in the large
$N$ limit at $a (T) = 1$. Thus the Hagedorn transition in the AdS string
theory can be studied by examining the
critical behavior of \DEFMo\ near $a =1 $. We study the matrix
model \DEFMo\ to all orders in the $1/N$ expansion and find that there
exists
 a double scaling limit
 \eqn\douSd{
 a-1 \to 0, \qquad N \to \infty , \qquad (a-1)N^{4 \ov 3} = {\rm
 finite}
 }
in which the transition is smoothed out. Expressed in the language of
the corresponding AdS string theory the limit is
 \eqn\dousSt{
 T - T_H \to 0, \qquad {l_{5p}  \ov R} \to 0, \qquad
(T-T_H) \le({R \ov l_{5p}} \ri)^{2} = {\rm finite}, \qquad
 l_{5p}^3 = {l_p^8 \ov R^5}
 }
where $l_{5p}$ is the five-dimensional Planck length in AdS$_5$. The
physical meaning of \douSd--\dousSt\ can be understood as follows.
At infinite $N$, there is a sharp phase transition at $T=T_H$. At large but
finite $N$ (or $R/l_{5p}$), the transition is smoothed out into a finite
region around $T_H$ whose width scales inversely with $N$ (or $R/l_{5p}$)
with some
critical exponent. The above double scaling
limit ensures that we stay within this region and decouples
nonessential physics.

In the double scaling limit \douSd--\dousSt, one finds
a simple physical picture
for the Hagedorn transition of the bulk string theory.
This picture is reminiscent of the Hawking-Page
transition~\refs{\hawkingpage,\witten} in the strong coupling
regime. We find that the Hagedorn divergence is resolved by
summing over two different
classical  {\it stringy} geometries. Since we are working
in the $l_s \to \infty$ limit \reime\ for the bulk string theory,
the  concept of geometry in terms of classical gravity is not
valid here. Nevertheless, in the large $N$ limit, classical
stringy geometry,  defined by exact worldsheet conformal field
theory, is still a valid concept.

We find that near the Hagedorn
transition (both above and below), the full partition function of
the theory can be written in a form
 \eqn\reWf{
 Z = Z_{T} + e^{- S_L} Z_L \ .
 }
$Z_T$ can be interpreted as the partition function for strings in
an AdS background with imaginary time-direction periodically
identified (we will call it thermal AdS), which contains a
Hagedorn divergence. $Z_L$ is the partition function of a
noncritical string theory which emerges only as the Hagedorn
temperature is approached. We find that it
has the same scaling as the unconventional branch of the
$\NN=1$ super-Liouville theory or a branched polymer.
$S_L$ can be interpreted as the difference in the classical string field
action
between the above two string backgrounds. When $T< T_H$,
$e^{-S_L}$ is exponentially small in large $N$ and the thermal AdS
dominates. At $T \approx T_H$, $S_L$ is of order one and the
two string backgrounds
are equally important\foot{More precisely, we mean $T-T_H \sim
O(N^{-2})$.}. In particular, the Hagedorn divergence in $Z_T$ is
cancelled by a similar divergence in $Z_L$. In this regime, the
theory does not have a (stringy) geometric interpretation. When $T
> T_H$, $e^{-S_L}$ becomes exponentially large and the noncritical string
background completely dominates. $Z_L$ is tachyonic below $T_H$
while the thermal AdS is tachyonic above $T_H$.

We also find some interesting non-renormalization properties for
$Z_T$ and $Z_L$. $Z_T$ contains only a genus one contribution with
all higher loop contributions vanishing. Below $T_H$, $Z_L$ also
does not receive perturbative corrections beyond one-loop. Above
$T_H$, $Z_L$ does receive perturbative contributions to all loops.

The plan of the paper is as follows. In section 2 we discuss
 the large $N$ Hagedorn transition of $\NN=4$ SYM theory at
infinite $N$. In section 3 we discuss the Hagedorn transition at
finite $N$ in Yang-Mills theory. We find that there exists a
double scaling limit in which the partition function smoothly
interpolates between the low and high temperature regimes. We
discuss the interpretation of Yang-Mills calculation in terms
of string theory in AdS.
We conclude with discussions of implications of our findings in
section~4. In Appendix A, we discuss a generalization of \DEFMo\
which includes both single and double trace terms. This arises
when one includes fundamental matter in the theory. We show that
the double scaling limit of the model is the same as that of the
single-trace unitary matrix model.

\newsec{Free large $N$ Yang-Mills theory on $S^3$}

In this section we  briefly review  the Hagedorn transition for a
free $SU(N)$ Yang-Mills theory (with adjoint matter) on $S^3$ in
the infinite $N$ limit~\refs{\sundborg,\aharony}. While the results
presented here are not new, our derivation of the phase transition is new.
The basic idea
is to integrate out all fields in the theory except for the zero
mode of the Polyakov loop. The partition function is then reduced
to a unitary matrix integral.

 Expanding all
fields in Yang-Mills theory in terms of harmonics on $S^3$, to the
lowest order in $g_{YM}$, the theory reduces to a quantum
mechanics problem of free harmonic oscillators
 \eqn\simlad{
\LL = \ha \sum_a \Tr \le[ (D_t M_a)^2  - \om_a^2 M_a^2 \ri]
 }
where $M_a$ are $N \times N$ Hermitian matrices and the sum is
over all field types and their Kaluza-Klein descendants on $S^3$.
$\om_a$ is the frequency for each mode. We will take the radius of
$S^3$ to be one. The covariant derivative in \simlad\ is given by
$$
D_t M_a = \p_t M_a - i [A, M_a]
$$
$A$ comes from the zero mode (i.e. the mode independent of
coordinates on $S^3$) of the time component of the gauge field and
is not dynamical. Its equation of motion imposes the constraint
that physical states must be $SU(N)$ singlets, i.e. Gauss law on
$S^3$. The partition function of the theory at finite temperature
can then be obtained by integrating out all fields in \simlad\ except for
$A$. Projecting into the singlet sector of the standard results for harmonic
oscillators one finds
 \eqn\resPar{\eqalign{
 Z & = \int DU \, \prod_a
 \le(\det_{adj} \le(1- \ep_a e^{-\beta \om_a} U \ri)\ri)^{-\ep_a}  \cr
 }}
where $U$ is an  $SU(N)$ matrix whose integration imposes the
singlet condition\foot{Note that we normalize the measure $DU$ so
that
$
\int D U = 1
$.}. It can also be interpreted as the Wilson line for $A$ around the
Euclidean time circle (Polyakov loop). $\det_{adj}$ denotes the
determinant in the adjoint representation and $\ep_a =  1$ ($-1$)
for bosonic (fermionic) $M_a$. Equation \resPar\ can be more
conveniently written as
 \eqn\rePae{
 Z =  \CC (\beta) \int DU \, \exp \le(\sum_{n=1}^\infty {w_n \ov n}
 \Tr U^n \Tr U^{-n} \ri)
 }
with
 \eqn\Defw{
 w_n = z_B (n \beta) +(-1)^{n+1} z_F (n\beta) \ .
 }
$z_B(\beta), z_F (\beta) $ are the single particle partition
function for the bosonic and fermionic sector respectively, e.g.
 \eqn\singP{
 z_B (\beta) = \sum_{a,{\rm bosons}} e^{-\beta \om_a},
\qquad
 z_F (\beta) = \sum_{a,{\rm fermions}} e^{-\beta \om_a} \ .
 }
The prefactor $C(\beta)$ in \rePae\ is given by
 \eqn\PrefC{
 \log \CC (\beta) = \beta E_0 - \sum_{n=1}^\infty {w_n \ov n} \ .
 }
$E_0$ is the zero-point energy of the theory on $S^3$ and the
second term arises because the group is $SU(N)$ rather than
$U(N)$. For free $\NN=4$ SYM theory on $S^3$, we
have~\refs{\sundborg,\aharony}
 \eqn\nequaf{
z_B =  {6x + 12 x^2 - 2x^3 \ov (1-x)^3}, \qquad
 z_F = {  16 x^{3 \ov 2} \ov (1-x)^3}, \qquad x = e^{-{\beta }} \
 .
 }

The large $N$ limit of the matrix integral \rePae\ can be
evaluated using a saddle point approximation~\refs{\bipz,\gw}.
Depending on the values of $w_n$, it could have a complicated
phase structure~\refs{\polaP}. For $\NN=4$ SYM, one finds from
\nequaf\ that parametrically
 \eqn\paraM{
{w_n \ov w_1} \ll 1 , \qquad n > 1
 }
and only two phases arise~\refs{\sundborg,\aharony} with the
critical temperature $T_H$ given by the solution of the equation
$w_1 (T_H) =1$. \paraM\ implies that one can approximate \rePae\
by the first term in the exponential
 \eqn\truni{
 Z \approx \int DU \, e^{w_1 \Tr U \Tr U^\dagger}
 }
In particular, since the critical behavior of \rePae\ near $w_1=1$
is controlled by the first term, one expects that the two models
\rePae\ and \truni\ should have the same critical behavior. Since
we are interested in the transition region, for the rest of the
paper we will focus on \truni.

The matrix model \truni\ can be solved to all orders by
introducing a Lagrange multiplier to eliminate the double trace
term in the exponential\foot{ I thank M. Douglas and V. Kazakov
for discussions on this point. A similar method was also used in
the context of double trace deformation of hermitian matrix models
by Klebanov and collaborators~\refs{\klebanov,\klebC}.}
 \eqn\themaT{\eqalign{
 Z(a) & = \int DU \, \exp \left(a \Tr U \Tr U^\dagger \right) \cr
 & = {1 \ov 2 \pi a} \int DU d \lam d \bar \lam \,\, \exp \left[
 -{1 \ov a} \lam \bar \lam + \lam \Tr U + \bar \lam \Tr U^\dagger
 \right] \ . \cr
 }}
Absorbing\foot{This turns the $SU(N)$ integral into an $U(N)$
integral.} the phase of $\lam$ into $U$ and letting
$$
|\lam| = \ha N g \
$$
we find that
 \eqn\thema{\eqalign{
 Z(a) & =  {N^2 \ov 2a}
\int_0^\infty g dg \, \exp \left( - {N^2 g^2 \ov 4a} + N^2 F (g)
  \ri) \cr
 }}
where $F (g)$ is given by the unitary matrix integral
 \eqn\defpaF{
  e^{N^2 F (g)} = \int D U \; \exp \le( \ha Ng (\Tr U + \Tr U^\dagger
 )\right) \ .
 }
Now $Z(a)$ reduces to an integral of the partition function of a
unitary matrix model weighted by a Gaussian factor over its
coupling constant.

The large $N$ expansion of matrix integral
\defpaF\ is well known~\refs{\gw,\goldsm,\periw} and the leading
order term is given by
 \eqn\DefFz{
 F(g) = \cases{{g^2 \ov 4} & $g  \leq  1$
\cr\cr
  g - \ha \log g - {3 \ov 4} & $g > 1 $ \cr
 }}
The discontinuity in the third derivative of $F(g)$ at $g=1$ is
the Gross-Witten large $N$ phase transition~\refs{\gw,\wadia}. The
$g < 1$ phase corresponds to
a saddle point of \defpaF\ at which
the eigenvalues $e^{i \th_i}, i=1, \cdots N$ of $U$ are
distributed around the whole unit circle. In particular at $g=0$,
the distribution is uniform. In the $g > 1$ phase, the eigenvalue
distribution develops a gap and does not cover the whole unit circle.
When $g \to \infty$ all eigenvalues are localized at $\th_i =0$.

The leading order result for $Z(a)$ is now simply given by the
saddle point of the $g$-integral \thema. The Gaussian weight
factor picks the eigenvalue distributions for different values of
$a$ from those of \defpaF. We introduce
 \eqn\DefQ{
Q (a,g) = -{g^2 \ov 4a} + F(g)
 }
 and look for local maximums of $Q$. We find that:

\item{1.} For $a < 1$, $Q(a,g)$ is a monotonically decreasing
function of $g$ with maximum at
 \eqn\MAxI{
g=0, \qquad Q(a,0) = 0
 }
 Expanding $Q(a,g)$ around $g=0$ and performing the
Gaussian integral we find that
 \eqn\Leadas{
 \log Z (a) = - \log (1-a) + \cdots
 }

\item{2.} For $a > 1$, $Q (a,0) = 0$ is now a local minimum. The
function has a maximum at
 \eqn\maxval{
g_0 =  {1 \ov 1 - w} > 1, \qquad w =\sqrt{1-{1 \ov a}}
 }
with
 \eqn\leadaL{\eqalign{
 \log Z (a) = N^2 Q(g_0) & = {N^2 \ov 2} \le({w  \ov 1 - w} +
 \log (1 - w) \ri)  + \cdots  \cr
 }}

\ndt\  For $\NN=4$ SYM, $a(T)$ is a monotonically increasing
function of $T$
 \eqn\DefA{
a(T) = w_1 =  {2 x (3-\sqrt{x}) \ov (1-\sqrt{x})^3}, \qquad x =
e^{-{\beta }}
 }
with the critical temperature at
 \eqn\HagT{
 w_1 (\beta_H) =1, \qquad
 \beta_H = 2 \log (2 + \sqrt{3}) \ .
 }
 There is no contribution of
$O(N^2)$ in the low temperature phase partition function \Leadas\
and the genus one term diverges as the critical temperature is
approached,
  \eqn\Hagd{
\log Z  \approx  -\log ( \beta-\beta_H)  + {\rm const}, \qquad T
\to T_H
  }
It follows that the density of states around $T_H$ is (which can
be found by a Laplace transform of \Leadas)
 \eqn\lowtEn{
 \Om (E)
 \approx   {\rm const} \; e^{\beta_H E} \le(1+  O(1/E^2) \ri)
 }
Note that \Hagd\ and \lowtEn\ are typical of Hagedorn behavior. In
fact they have exactly the same form as those in flat space string
theory with all spatial directions compactified (see e.g.
~\refs{\TanI,\BranV}). At order $O(N^2)$ there is a weakly first
order transition at $a =1$. The high temperature phase has
 nonzero genus zero contribution\foot{Note that although \leadaL\ is continuous
as $a \to 1$, the second derivative of $\log Z(a)$ (i.e. specific
heat) becomes singular.}.

In next section we will look at finite $N$ corrections to the
above results. We would like to understand how finite $N$
corrections resolve the singularity in \Hagd\ and smooth out the
transition. We conclude this section by making a few remarks on
the physical interpretation of~$U$ in \rePae.

\item{1.} $U$ can be understood as the temporal Wilson line for
$A$ around the time circle $\tau$ (Polyakov loop)
 \eqn\uniWil{
 U^n = P \exp \left(i \int_0^{n \beta}  A d \tau \ri)
 }
The theory has a $Z_N$ symmetry,
 \eqn\DiscS{
U \to e^{2 \pi i k \ov N} U , \qquad k \in \IZ
 }
coming from gauge transformations which are periodic up to an
element of the center $Z_N$. In terms of the eigenvalues $e^{i
\th_i}, i=1,\cdots N$ of $U$ the $Z_N$ symmetry corresponds to
discrete translations $\th_i \to \th_i + {2 \pi  k \ov N}$. In
fact  \rePae\ has a larger ``accidental $U(1)$ symmetry'', since
it was obtained from a free theory. The correlation functions of
the theory, however, do not have to respect this larger symmetry,
since the measure $DU$ for $SU(N)$ is only $Z_N$ invariant.

\item{2.} In the large $N$ limit, it is convenient to introduce
density of eigenvalues
 \eqn\chCor{
\rho (\theta) = {1 \ov N} \sum_{i=1}^N \delta(\theta - \th_i) ,
\qquad - \pi \leq  \th <  \pi
 }
Then $${1 \ov N} \Tr U^n =\rho_n = \int_{-\pi}^\pi {d \th } \,
\rho (\th) \,e^{i n \th}$$
 and  \rePae\ can be written as
 \eqn\patI{
 Z = \CC \int [D \rho]
 \, e^{- V [\rho]}
 }
where $V(\rho)$ has the form
 \eqn\Impote{\eqalign{
 V [\rho] & 
 = N^2 \sum_{n=1}^\infty {1 - w_n \ov n} \, |\rho_n|^2 \cr
 }}
$\rho_n$ can be interpreted in the dual AdS string theory  as the
$n$-th winding modes around the time circle. From \Impote\ and
\HagT, $\rho_1$ become massless at  the Hagedorn temperature and
tachyonic above $T_H$. The Hagedorn divergence \Hagd\ is precisely
due to that $\rho_1$ becoming massless~\refs{\sundborg,\aharony}.
In the low temperature
phase, the eigenvalue distribution is given by that of \defpaF\ at
$g=0$, which is
$$
\rho (\th) = {1 \ov 2 \pi} \ .
$$
In the high temperature phase, $\rho(\th)$ develops a gap.

\item{3.} The $Z_N$ symmetry \DiscS\ becomes an $U(1)$ in the
large $N$ limit, corresponding to
 \eqn\tranS{
\th \to \th + \al
 }
Under the transformation \tranS, $\rho_n$ transform as
$$
\rho_n \to \rho_n e^{- i n \al}
$$
It is tempting to interpret the $\th$-circle as the dual time
circle in AdS string theory. However, here $\apr = \infty$, it is
not clear how to define T-duality. The $U(1)$ symmetry of the
originial  time circle $\tau \to \tau + \ep$ is not manifest in
the matrix model. Due to the symmetry \tranS, given a saddle point
solution $\rho(\th)$ of \patI, $\rho (\th + \al)$ is a solution
for any $\al$. Thus there are a continuous family of saddle points
in the high temperature phase.

\newsec{Hagedorn transition at finite $N$}

In this section we would like to understand how finite $N$
corrections in Yang-Mills theory resolve the singular behavior in
the large $N$ expansion and  smooth out the transition. From the
AdS/CFT correspondence, this tells us how non-perturbative stringy
effects smooth out the singular Hagedorn behavior in perturbation
theory for the bulk string theory.

For this purpose, we need to work out the sub-leading corrections
to the matrix integral \themaT\ near the transition region (i.e.
$a \approx 1$). This can be obtained by expanding around the
saddle points of
  \eqn\thema{\eqalign{
 Z(a) & =  {N^2 \ov 2a} \int_0^\infty g dg \, \exp \left( - {N^2 g^2 \ov 4a} + N^2 F (g)
  \ri) \cr
  & =  {N^2 \ov 2a} \int_0^\infty g dg \, e^{N^2 Q}
  \cr
 }}
using the  large $N$ expansion of the unitary matrix model
 \eqn\defpaF{
  e^{N^2 F (g)} = \int D U \; \exp \le( \ha Ng (\Tr U + \Tr U^\dagger
 )\right) \ .
 }
We will first review the results for \defpaF.

\subsec{Unitary matrix model}

The asymptotic expansion of \defpaF\ in large $N$ is well
known~\refs{\goldsm,\periw}. Depending on the value of $g$, the
expansion can be divided into three different regions
 \eqn\moreXQ{\eqalign{
 N^2 F (g) \approx \cases{ {N^2 g^2 \ov 4} +
    {1 \ov 2 \pi }
    e^{-2 N f(g)} \sum_{n=1}^\infty {1 \ov N^{n}}
       F_n^{(1)}  & $g  <  1$ \cr\cr
     {N^2 g^2 \ov 4} +
     \sum_{n=0}^\infty N^{-{2 \ov 3} n} F_n^{(2)} & $g -1\sim O(N^{-{2 \ov 3}})$
    \cr\cr
     N^2 \le( g - \ha \log g - {3 \ov 4} \ri) +
    \sum_{n=0}^\infty N^{-2n} F_{2n}^{(3)} & $g > 1$ \cr
     }
     }}
where
 \eqn\Defg{
f(g) =  \log\left({1 \ov g} + \sqrt {{1 \ov g^2 } -1} \ri) - g
  \sqrt{{1 \ov g^2} -1}
 }
Explicit expressions for various series $F_n^{(1)}, F_n^{(2)},
F_n^{(3)} $ are not important for our discussion below and some
important features of \moreXQ\ are:

\item{1.} There are no perturbative corrections to $F(g)$ for
$g<1$ beyond the leading term. All corrections are
non-perturbative in $N$.

\item{2.} Sub-leading terms in the expansions for $g<1$ and $g>1$
become singular as $g \to 1$,
 \eqn\signB{
 F_n^{(1)} \sim {1 \ov (1-g)^{3n \ov 2}} , \qquad
 F_n^{(3)} \sim {1 \ov (g-1)^{3n }} \ .
 }
The asymptotic expansions in $1/N$ break down and are replaced by
that in intermediate region $|g-1| \sim N^{-{2 \ov 3}}$.

\item{3.} One can define a double scaling limit
 \eqn\doucU{
g = 1- N^{-{2 \ov 3}} t, \qquad N \to \infty
 }
which extracts the $F_0^{(2)}$ term from the expansion.
$F_0^{(2)}$ satisfies the equation
 \eqn\uniST{
 {d^2 \ov dt^2} F_0^{(2)}(t) = -f^2 (t)
 }
where $f(t)$ in turn satisfies the Painleve II equation
 \eqn\painle{
\ha f''(t) = f^3 + t f(t)
 }
Higher order terms $F_n^{(2)}$ can be shown to satisfy more
complicated  differential equations.
 When $|t| \gg 1$, $F_0^{(2)}$ has the following asymptotic
expansion
 \eqn\mquanii{
 F_0^{(2)} = \cases{ 
  {t^3 \ov 6}
 - {1 \ov 8} \log (-t) - {3 \ov 128 t^3}
 + {63 \ov 1024 t^6} + \cdots  & $-t \gg 1$ \cr\cr
  {1 \ov 2 \pi} e^{-  {4 \sqrt{2} \ov 3} t^{3 \ov 2}}
 \le( - {1 \ov 8 \sqrt{2} t^{3 \ov 2} } + {35 \ov 384 t^3} -
 {3745 \ov 18432 \sqrt{2} t^{9 \ov 2}} + \cdots \ri) & $ t \gg 1$ \cr
 }}
Note that $F_0^{(2)}$ is a smooth function of $t$ and interpolates
smoothly between $g>1$ ($t <0$) phase and $g<1$ ($t >0$) phase.

\item{4.} It was argued in~\refs{\kms} that $F_0^{(2)} (t)$
describes the full partition function of the type 0B theory in
$d=0$ dimension, i.e. pure 2-d supergravity. The parameter $t$ is
proportional to the cosmological constant $\mu$ in the
super-Liouville interaction.

\subsec{Critical behavior around $a \approx 1$ and a double
scaling limit}

We are interested in understanding the critical behavior of the
matrix model \thema\ near the transition point $a \approx 1$. We
emphasize that the integral \thema\ is manifestly finite for all
values of $a$. The Hagedorn divergence arises as a result of doing
the large $N$ expansion. To see how the Hagedorn divergence is
cancelled in the finite $N$ theory, it is convenient to split the
integral in \thema\ into
 \eqn\intsep{\eqalign{
 Z & = \int_0^1 dg  \, g \, e^{-{N^2 g^2 \ov 4} \le({1 \ov a}-1
 \ri) } + Z_1 + Z_2 \cr
 & =  {1 \ov 1-a} \le(1 - e^{-{N^2 \ov 4a} (1-a)} \ri)
 + Z_1 + Z_2
 }}
with
 \eqn\defoTH{\eqalign{
 & Z_1 = {N^2 \ov 2a} \int_0^{1} g dg \, e^{-{N^2 g^2 (1-a) \ov 4a} }
  \le(e^{\tilde F} -1 \ri) \cr
 & Z_2 = {N^2 \ov 2a} \int_{1}^\infty g dg \, e^{N^2 Q} \cr
 }}
where $\tilde F(g)$ is defined for $g < 1$
 \eqn\DEFFt{
 \tilde F = N^2 F(g) - {g^2 N^2 \ov 4}
 }
In \intsep\ we separated the integration for $g$ into $g<1$ and $g
>1$ regions motivated from \moreXQ. We further isolated from the
$g <1$ part of the integral the term ${1 \ov 1-a}$, which gives
rise to the Hagedorn divergence \Leadas, and a non-perturbative
term
 $$
- {1 \ov 1-a} e^{- N^2 f (a)} , \qquad f (a) = {1-a \ov 4a}
$$
which precisely cancels the divergence at $a=1$. This latter term
is not seen in perturbation theory. One can also readily check
that both $Z_1$ and $Z_2$ are convergent integrals for all $a$.
They can be evaluated by saddle point approximations. Notice that
$\tilde F$ defined in \DEFFt\ is exponentially small except near
$g=1$ (see \moreXQ\ and \Defg). When $a<1$, one finds that $Z_2$
is exponentially smaller than $Z_1$ and $Z_1$ is dominated by
 a saddle point which approaches $1$ as $a \to 1_-$. When $a >1$, $Z_1$ is
 exponentially smaller than $Z_2$ and $Z_2$ is
dominated by the saddle point \maxval\ which again approaches $1$
as $a \to 1_+$. It thus follows that the critical behaviors of the
theory around $a =1$ can be obtained by evaluating
\defoTH\ around the region $g \approx 1$.

To find the scaling region, let us first look at the limit of $a
\to 1$ from the high temperature side \leadaL. As $a \to 1_+$, we
find that
 \eqn\Aonep{\eqalign{
 & g_0 = 1 + \sqrt{a-1} + O(a-1), \cr
  &  \log Z(a) = N^2 \le( {a-1 \ov 4} + {1
\ov 3} (a-1)^{3 \ov 2} + O((a-1)^2) \ri) \cr
 }}
Although $\log Z(a)$ is continuous as $a \to 1_+$, its second
derivative with respect to $a$ (i.e. specific heat) becomes
singular in the limit. The leading non-analytic term $N^2 (a-1)^{3
\ov 2}$ in $\log Z(a)$ suggests a scaling $ a-1 \sim O(N^{-{4 \ov
3}})$. This also follows from $a-1 \approx (g_0-1)^2$ and the
scaling of the unitary model \doucU. This motivates us to define a
double scaling limit
 \eqn\Definvt{
1-a = N^{-{4 \ov 3}} s,  \qquad N \to \infty \ .
 }
With a change of variable
 \eqn\ChnV{
{ g} =  1 - N^{-{2 \ov 3}} t, \qquad
 }
the exponent in \thema\ can be written as
 \eqn\ForEx{\eqalign{
 N^2 Q & = {N^2 g^2 \ov 4} \le(1 -{1 \ov a} \ri) + F_0^{(2)} + N^{-{2 \ov
 3}} F_1^{(2)} + \cdots \cr
 & = - {1 \ov 4} N^{2 \ov 3} s + \ha st + F_0^{(2)} (t) + O(N^{-{2 \ov
 3}})  \cr
 }}
After some simple algebra, one finds that
 \eqn\finfZ{\eqalign{
 Z  & =  {N^{4 \ov 3} \ov s} \le(1 - e^{-{1 \ov 4} N^{2 \ov 3} s} \ri)
 + {N^{4 \ov 3}} e^{-{1 \ov 4} N^{2 \ov 3} s}
 \sum_{n=0}^\infty
 N^{-{2n \ov 3}} B_n (s) \cr
 &  = {N^{4 \ov 3} \ov s} \le(1 - e^{-{1 \ov 4} N^{2 \ov 3} s} \ri)
 + {N^{4 \ov 3}} e^{-{1 \ov 4} N^{2 \ov 3} s} B_0 (s) + \cdots \cr
 } }
where $B_n$ can be obtained from $F_n^{(2)}$.  The leading term
$B_0 (s)$ is  given by
 \eqn\DEFd{\eqalign{
 & B_0 (s) =  \ha \int_{-\infty}^{\infty} dt \, e^{\ha st } \le(e^{F_0^{(2)}} -\th(t) \ri)
 \cr}}
 where $\th(t)$ is the step function.
 From \DEFd\ and \mquanii\ $B_0 (s)$ is a completely smooth function of $s$ including $s=0$.
 Higher order terms $B_{n}, n>0$ are
 increasingly complicated and we have not looked at them in detail.
  Our procedure should guarantee that all of them are well defined
  and smooth functions of $s$. They are not relevant  in the
  double scaling limit \Definvt.

In the double scaling limit \Definvt, keeping only the leading
term in \finfZ, we have
 \eqn\AnaE{\eqalign{
 Z  & = {N^{4 \ov 3} \ov s} + {N^{4 \ov 3}} e^{-{1 \ov 4} N^{2 \ov 3} s}
 \tilde B_0 (s)   \cr
 }}
where we have introduced
 \eqn\DEFtB{
 \tilde  B_0 (s) = B_0 (s) - {1 \ov s} \ .
 }
Note that as $s \to 0$, $\tilde B_0$ is singular
 \eqn\singB{
 \tilde B_0 (s) \sim - {1 \ov s} + \cdots
 }
This singularity precisely cancels with that of the first term in
\AnaE. For $s< 0$, $\tilde B (s) $ can also be written as
 \eqn\Legd{
 \tilde B_0 (s)  =  \ha \int_{-\infty}^{\infty} dt \, e^{\ha st +
 F_0^{(2)} (t)}
 }
while for $s > 0$
 \eqn\explB{
 \tilde B_0 (s) = \ha \int_{-\infty}^{\infty} dt \, e^{\ha st } \le(e^{F_0^{(2)}} -1 \ri)
 }

One can easily generalize the above discussions to unitary models
with both single trace and double trace terms, e.g.
 \eqn\mixma{
Z (a, \mu) = \int d U \, \exp \le[a \Tr U \Tr U^\dagger + \ha N
(\mu \Tr U + \bar \mu \Tr U^\dagger ) \ri]
 }
where $U$ is a unitary matrix. \mixma\ is relevant when we include
$N_f$ matter fields in the fundamental representation of the gauge
group with $N_f/N$ to be finite\foot{The actual matrix model
resulting by integrating out all fields except for the zero modes
of the Wilson loop is more complicated. But again one can argue
that \mixma\ is enough for understanding the critical behavior of
the theory.}. One can show that the critical behaviors of \mixma\
are in fact the same as those of the single-trace unitary matrix
models. This implies there is no Hagedorn type transition for
\mixma~\refs{\SchnitzerQT}. Some details can be found in Appendix
A.

\subsec{AdS interpretation}

The partition function $Z$ \AnaE\ has the following form
 \eqn\reWf{
 Z = Z_{T} + e^{- S_L} Z_L
 }
with
 \eqn\varPa{\eqalign{
 \log Z_T & = -\log s + {\rm const} \cr
  \log Z_L & = \log \tilde B_0 (s) + {\rm const} \cr
  }}
  and
 \eqn\acCl{
 S_L = N^{2 \ov 3} s = N^2 (1-a(T)) \approx N^2 a'(T_H) (T - T_H)
 }
$Z_T$ is due to the saddle at $g=0$ (uniform distribution of
matrix eigenvalues) and  $Z_L$ comes from  $Z_1 + Z_2$  at $g
\approx 1$.

We would like to interpret equation \reWf\ in terms of contributions from
two different {\it
classical stringy geometries} in the corresponding
AdS string theory. The free Yang-Mills theory we are working with
corresponds to the
 $l_s \to \infty$ limit \specL\ for the bulk string theory. Thus
the  concept of geometry in terms of classical gravity is not
valid here. Nevertheless, since we are working in the large $N$
limit, classical stringy geometry, as defined by conformally
invariant sigma-model on the worldsheet, is still a valid concept.
In particular, saddle points (local maxima) in the matrix model
\rePae\ should correspond to  worldsheet conformal field theories.
The saddle points of \rePae\ in turn coincide with those of
\thema\ in the scaling region \Definvt\ we are interested in here.
Thus we shall interpret \reWf\ as summing over two classical
string backgrounds. $S_L$ can be interpreted as the difference in
the classical string field action for two backgrounds and $Z_T, Z_L$ are
partition functions around each background. Clearly $\log Z_T$
should describe string theory in the thermal AdS geometry, i.e.
AdS with the time direction periodically identified. The precise
string theory interpretation of $Z_L$ is not clear to us. One
possible identification is that $\log Z_L$ describes the {\it
unconventional branch} of the super-Liouville theory with $s$
identified as the cosmological constant. In the next subsection we
will also mention other possibilities.

The physical picture reflected from \reWf\ can be summarized as
follows. We first divide our discussions into three regions and then comment
on some general aspects.

\item{1.} $T < T_H$, $s >0$ but not too close to zero. In this
region the $Z_T$  (thermal AdS) dominates and $Z_L$ is
exponentially suppressed. Note that $Z_T$ does not contain
perturbative corrections beyond one loop. Strictly speaking, $Z_L$
does not correspond to a local maximum of the full matrix model
\thema. Nevertheless, we find it natural to interpret it as
corresponding to a conformal field theory on the string theory
side. This theory appears to be tachyonic due to the negative sign
in equation \singB. For $s \gg 1$ we can evaluate \explB\ as an
asymptotic series in $1/s$. Surprisingly, we find that the
asymptotic series terminate beyond the Gaussian integration
 \eqn\aeqod{
 \log \tilde B_0 = {s^3 \ov 192} - {5 \ov 2} \log s + C' + i \pi
 + \le({\rm terms \;\; nonperturbative \; \; in \;\;} {1 \ov s}
 \ri)
 }
 where $C'$ is a real constant.
This implies that $\log Z_L$ does not receive perturbative
corrections beyond one-loop.
The imaginary term $i \pi$ in \aeqod\ again indicates that the
theory contains tachyonic modes. More precisely, it has the
behavior of a complex tachyon in $0$-Euclidean dimension.

\item{2.} The transition region: $s \sim O(N^{-{2 \ov 3}})$, i.e.
$T-T_H \sim O(N^{-2})$. In this region, the two terms in \reWf\
are of comparable strength. Since two backgrounds contribute
equally in this regime, the full string theory does not have a
well defined (stringy) geometric interpretation. The $1/s$
Hagedorn divergence of $Z_T$ is cancelled by a similar divergent
term in $Z_L$ (see \singB). We remarked below \Impote\ that the
Hagedorn divergence in $Z_T$ can be understood as $\rho_1$
becoming massless. What is the interpretation of the $1/s$
singularity for $Z_L$? There are two possible interpretations. One
is that \singB\ arises from the volume factor of the Liouville
theory, i.e. the theory becomes effectively noncompact when $s \to
0$\foot{Of course, the numerical constant multiplying the volume
factor has to match \singB.}. A second interpretation is that
$Z_L$ develops a $0$-dimensional complex massless mode (let us
call it $\sig_1$) as $s \to 0$, given that the theory is tachyonic
for $s>0$ and not tachyonic for $s<0$. If true, it means that for
$s>0$, $\sig_1$ is tachyonic, while for $s<0$, $\rho_1$ becomes
tachyonic.

\item{3.} $T>T_H$, $s < 0$ but not too close to zero. The saddle
at $g \approx 1$ becomes the local maximum of the full integral.
$Z_L$ is exponentially large and dominates.  $g=0$ is no longer a
local maximum and $Z_T$ is not physically meaningful\foot{Since we
have omitted terms in \finfZ\ much bigger than $Z_T$.}. When $-s
\gg 1$, \Legd\  can be evaluated as asymptotic expansions in $1/s$
using \mquanii\
  \eqn\AnoSr{
\log \tilde B_0  =  {(-s)^{3 \ov 2} \ov 3} +
  C - {5 \ov 16} \log (-s)  +{35 \ov 96 (-s)^{3 \ov 2}} + {245 \ov 256
 (-s)^6} + O\le({1 \ov (-s)^{9 \ov 2}}\ri) \ .
 }
Treating $(-s)^{- {3\ov 4}}$ as the effective string coupling
constant, $\log Z_L$ contains perturbative corrections to all
orders.

\item{4.} $S_L$  \acCl, which we computed explicitly using the dual Yang-Mills
theory, can be interpreted to be the difference
of the classical string field action of two backgrounds. $S_L$
cannot be computed in
a first quantized formalism with strings moving in a fixed
background. In a second quantized formalism, like string field
theory, it is natural to expect $S_L$ to be computable.

\item{5.} Our discussion above was restricted to the region $T-T_H \sim
O(N^{-{4 \ov 3}})$. Going beyond this region requires understanding
the full matrix model \rePae, which is a rather nontrivil task.
We have studied the truncated model \themaT\ to all orders in $N$
in detail for general $a$. We find that for $T < T_H$, there are no
perturbative  corrections to the partition function of the thermal
AdS background beyond one-loop. We expect this to be true for the
full matrix model \rePae, given the quadratic nature of the action.
We also find that in \themaT\ there are non-perturbative corrections of
the form
 \eqn\Dins{
e^{-N \log N}
 }
and
 \eqn\Geom{
e^{-N^2 f (T)}
 }
\Geom\ again suggests contribution of another geometry.
 However, away from the scaling region, \Geom\  is subdominant compared
 to \Dins, so it is not clear whether it has an unambiguous meaning.
While it is tempting to interpret \Dins\ as due to D-instantons,
it is not clear to us how to understand the $\log N$ factor.

\subsec{The unconventional branch of super-Liouville theory}

It is a natural question to ask whether $\log \tilde B_0$ has an
interpretation in terms of a non-critical string theory\foot{This
subsection grew out of discussions with I. Klebanov and J. Maldacena.}.
The scaling behavior of \AnoSr\ corresponds to a string
susceptibility exponent given by
 \eqn\susbil{
 \ga_{st} = \ha
 }
Here we discuss two possible, perhaps related, interpretations.

A well known phase of random matrix models which has the value of
exponent \susbil\ is a branched polymer phase (see e.g.~\refs{\ambj}).
The branched polymer interpretation seems to fit with the picture
that in the high temperature (deconfinement) phase, quarks and gluons
are ``liberated'', and
the continuous Riemann surface description breaks down~\refs{\atick}.
The
scaling behavior \susbil\ also appeared in the double trace
deformations of Hermitian matrix
models~\refs{\das,\AlvarezGaumeZI,\KorchemskyTT}.
Based on Feynman diagrams generated by the double trace term, it was
argued in~\refs{\das} that the matrix model describes a branched
polymer phase, when the
coefficient of the double trace term is sufficiently large.
 In our case, the story is less clear since we do
not have a good geometric picture of how discretized worldsheets
arise in a unitary matrix model. It is also not clear to us how to
interpret the scaling behavior of \aeqod\ for $s >0$ from this
point of view.

We will now offer an alternative interpretation of \AnoSr, again
drawing inspiration from the analogous questions in the context of
double trace deformations of Hermitian matrix models.
In~\refs{\das} it was found that a new phase appears when the
coefficient of the double trace term takes a particular value.
Klebanov and Hashimoto~\refs{\klebanov,\klebC} gave an interesting
interpretation of this new phase. They observed that the scaling
behaviors of the new critical points in the presence of double
trace deformations can be explained by changing the branch of the
Liouville dressing. Our discussion of the double scaling limit,
and in particular, the expressions \Legd\ and \explB\ are rather
similar to those obtained there. Given that $F_0^{(2)}$ in \Legd\
and \explB\ is identified with the conventional branch of the
$\NN=1$ super-Liouville theory~\refs{\kms}, it seems natural to
identify $\log \tilde B_0$ with the other branch of the
super-Liouville theory. We will now check that the scaling in
\AnoSr\ is indeed consistent with this proposal with $(-s)$
identified with the Liouville cosmological constant. In Appendix A
we study a double trace deformation of the simplest single trace
unitary model and show that its critical behavior is the same as
the single trace case except for the case we discussed in previous subsections.

The super-Liouville action can be written as (we take $\apr =2$)
 \eqn\lag{\eqalign{
  S & = \frac{1}{2\pi}\int\!d^2 \sig d^2\theta\,
    \Bigl[ D\Phi \bar D\Phi + 2 i\mu_0 e^{b\Phi}\Bigr]\ \cr
  & =  {1 \ov 2 \pi} \int d^2 \sig \, \left[
  \p \phi \bar \p \phi + {1 \ov 4} Q R \phi + \psi \bar \p \psi
+ \bar \psi \p \bar \psi + \mu^2 b^2  e^{2 b \phi}
 + 2 i \mu_0 b^2 \psi \bar \psi e^{ b \phi} \right]
 \cr
 }}
with the central charge given by
$$
 \hat c_L = 1 + 2 Q^2
 $$
 and $b$ satisfying
  \eqn\DefBV{
 Q = b + {1 \ov b}
  }
For pure supergravity we need $\hat c_L = 10$ which leads to
  \eqn\posiB{
 Q = {3 \ov \sqrt{2}}
  }
There are two solutions to \DefBV,
 \eqn\negaB{
 b_- = {1 \ov \sqrt{2}}, \qquad  b_+ = \sqrt{2}
 }
where $b_-$ is the standard branch which satisfies the Seiberg
bound. $b_+$ branch does not correspond to a local operator.

The dependence of \lag\ on $\mu_0$ can  be obtained by taking
$$
\Phi \to \Phi - {1 \ov b} \ln \mu_0
$$
and we find the free energy on a genus $h$ surface is
$$
F_h \sim \mu_0^{\chi {Q \ov 2b}}
$$
For the  branch $b_-$ we have
 \eqn\Scalu{
F_h \sim \mu_0^{{3 \ov 2} (2-2h)} 
 }
The scaling agrees with that of the first line of \mquanii\ if we
identify $\mu_0 > 0$ with $-t$. For $\mu_0 < 0$, $F_h$ are
identically zero since there are only non-perturbative corrections
in the second line of \mquanii. For the other branch $b_+$ we have
 \eqn\TheoB{
F_h \sim \mu_0^{{3 \ov 4} (2-2h)} \ .
 }
This agrees with \AnoSr\ if we identify $-s \propto \mu_0 >0 $. It
seems natural to identify \aeqod\ with the same theory with $\mu_0
< 0$. However, the $s^3$ term in \aeqod\ does not seem to agree
with the scaling of \TheoB\ for $h=0$. There are several
possibilities for the disagreement\foot{It was also pointed out to us
by I. Klebanov that \aeqod\ has the same scaling as
the $c=-2$ matrix model.}. First notice that $s^3$ term
is analytic in $a-1$, so this term might not be universal and does
not have a Liouville interpretation. The second is that the
scaling \TheoB\ does not hold for $\mu_0 <0$ due to certain
subtleties.

It would be very interesting to have a more complete and precise
string theory identification of \Legd\ and \explB.

\newsec{Discussion and conclusions}

In this paper we investigated non-perturbative aspects of the
Hagedorn transition for IIB string theory in an anti-de Sitter
spacetime in the limit that the string length goes to infinity. We
find that as one approaches the Hagedorn temperature perturbative
string theory  breaks down and a noncompact Liouville direction
appears to open up\foot{In this section we will assume the
Liouville interpretation.}. In the double scaling limit $T- T_H
\sim O(N^{-{4 \ov 3}}), \;\; N \to \infty$, the full partition
function can be written as a superposition of those from the
thermal AdS and the Liouville background, weighed by their
relative classical action. The Hagedorn singularity of
perturbative string theory around thermal AdS is cancelled by a
similar divergence in the Liouville theory~\foot{This is also
reminiscent of the discussion in~\refs{\malet} where unitarity of
the AdS black hole background is restored by summing over other
geometries.}. Our discussions here apply  to a variety of
Yang-Mills theories or their string theory dual which have the
same critical behavior as the double trace unitary matrix model
\themaT.

The summing-over-geometry behavior we see here is rather similar
to that of the Hawking-Page transition~\refs{\hawkingpage,\witten}
in the strong coupling. There are some differences.
One difference is that we are summing over conformal field
theories instead of classical gravity backgrounds. Another
important difference is that in our case at a given temperature
only one background is stable. In the strong coupling limit, the
Hawking-Page transition appears at a temperature $T_{c} \sim
O(1)$, much lower than the Hagedorn temperature at $T_H \sim
O(\lam^{1\ov 4})$. When extrapolated to the strong coupling, the
phase transition here may become the Hawking-Page transition or
describe the transition between a meta-stable thermal AdS  to an
AdS black hole at $T_H$.

It is somewhat suprising and interesting that near the transition
point, the bulk theory can be described by a Liouville theory.
This is reminiscent of the appearance of long throats in
other singular CFTs like the conifold~\refs{\vafao,\witt,\kuta}.

We also found some interesting non-renormalization properties.
At temperatures below the Hagedorn temperature,
the string partition function around the
thermal AdS background appears to receive only one-loop contribution.
Equation \aeqod\ also does not receive corrections beyond one loop.
It would be interesting to understand these properties better.

Given the universal presence of the Hagedorn behavior in
perturbative string theories, the lessons we learned here might
serve as a useful guide for probing non-perturbative stringy
effects in other string theories including those in asymptotically
flat spacetime. For example, it is interesting to check whether
scaling behavior exists in type II or Heterotic theory in flat
spacetime as one approaches the Hagedorn temperature, i.e. we would like
to know whether leading
order Hagedorn divergences at higher genera have the form
 \eqn\HagD{
 \log Z \sim -\log (\beta - \beta_H) + \sum_{n=1}^\infty g_s^{2n}
 {1 \ov (\beta - \beta_H)^{\ga n}} ,
 }
with $\ga$ the critical exponent. We hope to come back to this question later.

\bigskip
\noindent{\bf Acknowledgments}

We would like to thank M.~Douglas, G.~Festuccia, V.~Kazakov,
I.~Klebanov J.~Maldacena, S.~Minwalla, J.~Minahan, G.~Semenoff,
A.~Sen, S.~Shenker, W.~Taylor, M.~Van Raamsdonk, B.~Zwiebach for
very useful discussions and especially J.~Minahan,  B.~Zwiebach
for very valuable help. This work is supported in part by funds
provided by the U.S. Department of Energy (D.O.E) under
cooperative research agreement \#DF-FC02-94ER40818.

\appendix{A}{Mixed unitary matrix model}

Here we consider the double scaling limit of the following unitary
matrix model
$$
Z (a, \mu) = \int d U \, \exp \le[a \Tr U \Tr U^\dagger + \ha N
(\mu \Tr U + \bar \mu \Tr U^\dagger ) \ri]
$$
which arises if one introduces fundamental
matter~\refs{\HallinKM,\SchnitzerQT}. Our discussion follows that
of~\refs{\klebC}. Introducing the Lagrange multiplier we find that
 \eqn\Prieq{\eqalign{
 Z (a, \mu) & = {N^2 \ov 8 \pi a} \int dU d \lam d \bar \lam \,\, \exp \left[
 -{N^2 \ov 4 a} (\lam - \mu) (\bar \lam - \bar \mu) + \ha N \le(\lam  \Tr U +
 \bar \lam
  \Tr U^\dagger \ri)  \right] \cr
 &= {N^2 \ov 4 \pi a}  \int_0^\infty g d g  \int_{-\pi}^\pi d \th \;
 \exp \left(-{N^2 \ov 4 a} \le(g^2 - g (\mu e^{-i \th} + \bar \mu
 e^{i \th} ) + \mu^2 \ri)
  + N^2 F (g) \ri) \cr
 & = {N^2 \ov 2 a}  \int_0^\infty g d g  \, e^{N^2 Q}
 }}
where
$$
N^2 Q =\log I_0 \le({N^2 g |\mu| \ov 2a}\ri) - {N^2 \ov 4a} (|\mu|^2 +
g^2) + N^2 F(g)
$$
and we have used
$$
 e^{N^2 F (g)} = \int d U \; \exp \le( \ha N g (\Tr U + \Tr U^\dagger
 )\right)
 $$
and
$$
\int_{-\pi}^\pi {d \th \ov 2 \pi} \; e^{z \cos \th} = I_0 (z)
$$
Note that we have introduced $g=|\lam| $
and in the second line of \Prieq\  we absorbed the phase of $\mu$
into the integration of $\th$.
Below for notational simplicity we will denote $|\mu|$ simply as $\mu$.
Thus $\mu >0$.

Note that
$$
\log I_0 \le({N^2 g \mu \ov 2a}\ri) ={N^2 g \mu \ov 2a} - \ha \log
 \le({\pi N^2 g \mu \ov a}\ri) - { a \ov 4 N^2 g \mu} + O(N^{-4})
$$

We will evaluate \Prieq\ using the saddle point approximation. We
will look at the leading order terms in $Q$. Using \moreXQ\ we find that

\item{1.} When $a<1$ and $\mu < \mu_0 = {1-a }$, $Q$ has a maximum
at
$$
g_0 = {\mu \ov \mu_0} < 1, \qquad Q (g_0) = {\mu^2 \ov 4(1-a)} =
{\mu^2 \ov \mu_0^2}{1-a \ov 4}
$$

\item{2.} When $a<1$ and $\mu > \mu_0$ or $a>1$, $Q$ has a maximum
at
$$
g_0 = a+ {\mu \ov 2} + \sqrt{(a+\ha\mu)^2 -a}, \qquad
 Q(g_0) = \ha \le(g_0 +{\mu \ov 2a} g_0 -1 -\log g_0 \ri) -{\mu^2
 \ov 4a}
$$

\ndt
 Thus it follows from the above that in the $a-\mu$ plane,
below the line $a+\mu=1$ (note  $\mu > 0$), the system is in the
phase in which the distribution of eigenvalues of $U$ has no gap
on the unit circle, while above the line $a+\mu=1$, the system
develops a gap in the distribution of eigenvalues of $U$.

The behavior of the system as the critical line is approached can
be obtained as follows. It is clear that as far as $\mu \neq 0$,
the saddle points of both phases approach $g_0=1$ as the critical
line is approached~\foot{This is different from $\mu=0$ case, in
which the saddle point is always at $g=0$ for $a <1$.}. More
explicitly, from the above equations one can check that from both
sides
$$
g_0 = 1+ {\Delta \ov 1-a} + O(\Delta^2), \qquad \Delta = \mu -
\mu_0 \ll 1
$$
Motivated from the scaling behavior of the unitary matrix model
\doucU, we  let
$$
\mu = \mu_0 (1- N^{-{2 \ov 3}} s), \qquad g = 1- N^{-{2 \ov 3}} t
$$
and
$$
N^2 Q = K_0 - {N^{2 \ov 3} \ov 4} {1-a \ov a} \le(t - {s } \ri)^2
+  F_0^{(2)} (t) + O(N^{-{2 \ov 3}})
$$
with
$$
K_0= {N^2 \ov 4} (1-a) + {N^{4 \ov 3} s (1-a) \ov 2} + {N^{2 \ov
3} \ov 4} {s^2 ( 1-a)} - \ha \log \le({N^2 \pi (1-a) \ov a}\ri)
$$
Plugging the above expressions into \Prieq\ we find that
$$
Z ={N^{4 \ov 3} \ov 2a} e^{K_0} \int_{-\infty}^\infty dt \,
 e^{- {N^{2 \ov 3} \ov 4} {1-a \ov a} \le(t - {s }
\ri)^2 +  F_0^{(2)} (t) } \le (1 + O(N^{-{2 \ov 3}}) \ri)
$$
We thus find that
$$
\log Z = \tilde K_0 + F_0^{(2)} \le(s  \ri) + O(N^{-{2 \ov 3}})
$$
with $\tilde K_0$ non-universal terms
$$
\tilde K_0 = {N^2 \ov 4} (1-a) + {N^{4 \ov 3} s (1-a) \ov 2} +
{N^{2 \ov 3} \ov 4} {s^2 ( 1-a)} - \log (1-a)
$$
Thus we find that with any nonzero $\mu$, the double trace
deformation of the unitary matrix model has the same critical
behavior as that of the unitary matrix model up to non-universal
terms. The case for $\mu=0$ needs separate treatment, which is discussed
in the main text.

 \listrefs
\end
\end